\documentclass[a4paper,11pt]{article}
\usepackage{jheppub} 
\usepackage{lineno}

\usepackage{cleveref}              
\usepackage{tikz}
\usepackage{setspace}
\usetikzlibrary{arrows,matrix,positioning}
\allowdisplaybreaks

\usepackage{bm}

\usepackage{slashed}

\def\be{\begin{equation}}
\def\ee{\end{equation}}
\def\bea{\begin{eqnarray}}
\def\eea{\end{eqnarray}}
\newcommand{\beq}{\begin{equation}}
\newcommand{\eeq}{\end{equation}}
\def\beal{\begin{equation}\begin{aligned}}
\def\eeal{\end{aligned}\end{equation}}

\newcommand{\dd}{{\mathrm{d}}}

\title{On-shell representation and further instances of the 2-split behavior of amplitudes}

\author[a,b]{Thales Azevedo,}
\author[b,c]{Humberto Gomez}
\author[b]{and Renann Lipinski Jusinskas}
\affiliation[a]{Instituto de F\'isica, Universidade Federal do Rio de Janeiro,\\
  Av. Athos da Silveira Ramos 149, 21941-972, Rio de Janeiro -- Brazil}
\affiliation[b]{Institute of Physics of the Czech Academy of Sciences \& CEICO,\\
  Na Slovance 2, 182 21, Prague -- Czech Republic}
\affiliation[c]{Facultad de Ciencias Basicas, Universidad Santiago de Cali,\\
Calle 5 Nº
62-00 Barrio Pampalinda, Cali, Valle -- Colombia}

\emailAdd{thales@if.ufrj.br}
\emailAdd{humgomzu@gmail.com}
\emailAdd{renannlj@fzu.cz}

\abstract{ The newly discovered  splitting behavior of tree-level
scattering amplitudes of particles and strings has been expressed in terms of currents containing one off-shell leg. In this work, we explain how to obtain on-shell representations of the split amplitudes in different theories. Furthermore, we show that this 2-split behavior is also verified in gauge and gravity theories involving higher-dimensional operators, thereby providing additional evidence to its universal character. As a byproduct, we also generalize the transmuting operators  to amplitudes in higher-derivative theories.}

\begin{document}
\maketitle
\flushbottom

\section{Introduction}

Since the turn of the century, the study of scattering amplitudes has transcended the traditional Feynman-diagrammatic intuition, in particular with the identification of novel/hidden structures and several remarkable properties. Among the more recent ones, two interrelated concepts stand out: hidden zeros, smooth splitting (3-split) and the so called 2-split processes \cite{Arkani-Hamed:2019vag,Cachazo:2021wsz,Arkani-Hamed:2023swr,Cao:2024gln,Cao:2024qpp,Arkani-Hamed:2024fyd,Guevara:2024nxd,Feng:2025ofq,Zhou:2025tvq,Feng:2025dci}. The former concerns the discovery that tree-level amplitudes vanish on special sub-loci of kinematic space, even though no obvious factorization channel is present.  These hidden zeros impose non-trivial constraints on the analytic structure of the amplitude and hint at the existence of deeper, universal factorization properties. For example, in the colored scalars of $\text{Tr}(\phi^3)$ theory one finds that when certain non planar invariants are taken to zero, the amplitude vanishes. Perhaps more significantly, near those loci the amplitude factorizes in a manner compatible with the 2 and 3-split descriptions \cite{Arkani-Hamed:2023swr} --- see also \cite{Cao:2024gln,Bartsch:2024amu,Zhou:2025tvq}. In essence, when a certain kinematic data set  is constrained to vanish, a scattering amplitude may factorize in a striking way into the product of two lower-point amputated currents. This phenomenon has been identified across a broad class of field theories (such as the bi-adjoint $\phi^3$ model, the non‐linear sigma model, Yang-Mills theory,  gravity) and even in string theory \cite{Cachazo:2021wsz,Cao:2024gln,Cao:2024qpp}.

Following the identification of hidden zeros, the 2-split phenomenon was systematically studied in the work by Cao, Dong, He and collaborators. In \cite{Cao:2024gln,Cao:2024qpp}, it was shown that when one imposes the vanishing of a subset of kinematic invariants, the scattering equations of the CHY framework or the Koba--Nielsen factors in the string integrand cleanly split into two independent sectors, hence the term ``2-split'' behaviour. Schematically,
$$
{\cal A}_n(1,2,\dots,n) \longrightarrow {\cal J}^{L}(i, A, j, \kappa)\times{\cal J}^{R}(1,\ldots,i,j, \ldots, {\kappa'}),
$$
where the left and right amputated currents are lower-point building blocks constructed in the same theory or in a related one via transmuting operators \cite{Cheung:2017ems}.

Crucially, Feng, Zhang, and Zhou \cite{Feng:2025ofq} have further shown that the 2-split behavior is deeply tied to hidden zeros. By employing a BCFW recursion (or its modifications for non-standard theories) one can derive the factorization and the vanishing loci together.

The 2-split factorization can be seen as a property of the integrand in the CHY representation, where the splitting of the scattering equations corresponds to separate integration domains that yield the product of two currents. The universality of the 2-split phenomenon extends across gauge theory and gravity, suggesting that the structure is a deep aspect of the S-matrix rather than a peculiarity of any individual model.
Additionally, the vanishing loci provide a geometric handle in the kinematic space. They delineate subspaces where the amplitude simplifies radically, offering a fresh path to recursion or factorization outside traditional collinear or multi-particle poles.

This interplay between  hidden zeros, 2-split factorization, and transmuting operators opens new avenues for amplitude computation. Towards this goal, it would be interesting to recast the 2-split in terms of physical quantities, amplitudes themselves instead of gauge-dependent amputated currents. Indeed, as discussed recently in \cite{Feng:2025ofq}, the currents appearing in the 2-splits of Yang‒Mills  and gravity amplitudes are sensitive to gauge choices. Moreover,  CHY and (amputated) Berends--Giele currents do not agree in general. It would be better to express the splittings in terms of gauge-invariant objects.

In this paper, we contribute to the understanding of the 2-split behavior in two manners. First, we show that the splitting is also verified in theories with higher-derivative kinetic terms, such as $R^2$ gravity or the $(DF)^2$ theory of \cite{Johansson:2017srf}. Second, we explain how   to obtain on-shell representations of the split amplitudes.
As a byproduct,  we also generalize the transmuting operators introduced in \cite{Cheung:2017ems} to amplitudes in higher-derivative theories.

\section{Review of the universal splitting behavior}\label{sec:review}




In this section, we briefly review the splitting behavior of the main CHY ingredients, namely the measure, the Parke-Taylor factor and the (reduced) Pfaffian.

In the CHY representation, an amplitude with $n$ external massless states is written as an integration over a $n$-punctured Riemann sphere,
\begin{equation} \label{eq:CHYamplitude}
    \mathcal{A}_n = \int_{{\mathbb{C}}^n} \dd\mu_n \,I_n \tilde{I}_n.
\end{equation}
The CHY measure \cite{Cachazo:2013hca,Cachazo:2013iea}, is given by
\begin{equation}
    \dd\mu_n= \frac{\dd^n z}{\textrm{vol SL}(2,\mathbb{C})} z_{i,j} z_{j,k} z_{k,i}{\prod_{c\neq i,j,k}} \delta (S_c),
    \label{equationmeasure}
\end{equation}
where $z_c$ denotes the coordinates of the punctures,  $z_{c,d} = z_c - z_d$, and $S_c$ is defined as
\begin{equation}\label{eq:SE}
    S_c= \sum_{d\neq c} \frac{s_{c,d}}{z_{c,d}}.
\end{equation} 
We have introduced the shorthand notation,
\begin{equation}\label{eq:dotproduct}
s_{a,b} = 2\, p_a \cdot p_b \, ,
\qquad
s_{a_1,a_2,\ldots,a_q}
=
2 \sum_{\substack{m,n=1 \atop m<n}}^{q}
p_{a_m} \cdot p_{a_n}  .
\end{equation}
For massless external legs, $p_a^2 =p_b^2= 0$, this definition coincides with the usual Mandelstam invariants, namely,
\begin{equation}
    s_{a,b} = 2\, p_a \cdot p_b = (p_a + p_b)^2 \, .
\end{equation}
However, when off-shell external legs are present, e.g. $p_\kappa^2 \neq 0$, the definition in Eq.~\eqref{eq:dotproduct} no longer matches the standard Mandelstam invariants,
\begin{equation}
    s_{a,\kappa} = 2\, p_a \cdot p_\kappa = (p_a + p_\kappa)^2 - p_\kappa^2\, .
\end{equation}
In this work, we adopt the dot-product notation of Eq.~\eqref{eq:dotproduct} purely for convenience.

The measure is independent of the choice of reference labels ${i,j,k}$.  In addition, we can use the automorphism group of the sphere, $\textrm{SL}(2,\mathbb{C})$,  to fix the coordinates of three punctures. For convenience we will fix the coordinates of the reference labels $\mathcal{I}=\{i,j,k\}$, with $i<j$ and $k=n$, such that the gauge fixed measure is recast as
\begin{equation}
    \dd\mu_n= \Delta(i,j,k)^2 {\prod_{c\neq i,j,k}} dz_c \, \delta (S_c),
    \label{eq:measure}
\end{equation}
with $\Delta(i,j,k)=z_{i,j}z_{j,k}z_{k,i}$. The  half-integrands $I_n,\tilde{I}_n$ are theory dependent and usually involve  other relevant kinematic data.

\subsection{The splitting of the measure}

The  splitting behavior analyzed  in \cite{Cao:2024gln,Cao:2024qpp} is  observed when separating the $n-3$ punctures to be integrated into two disjoint sets $A$ and $B$ such that  $A\cup B = \{1,2,\dots,n\}\setminus\mathcal{I}$, and then imposing the conditions
\begin{equation}
    s_{a,b} = 0 \qquad \forall \, a\in A, b\in B\,.
    \label{eq:main}
\end{equation}
Under the splitting conditions \eqref{eq:main}, the scattering equations $S_c=0$ take a suggestive form. Let us consider $S_a$ with $a\in A$, and $S_b$ with $b \in B$. They are respectively given by \begin{subequations} \label{eq:SEsplitting-on}
\begin{align}
S_{a}&= \frac{s_{a,i}}{z_{a,i}}+\frac{s_{a,j}}{z_{a,j}}+\frac{s_{a,k}}{z_{a,k}}+\sum_{\tilde{a}\in A\setminus\{a\}}\frac{s_{a,\tilde{a}}}{z_{a,\tilde{a}}},\\
S_{b}&=\frac{s_{b,i}}{z_{b,i}}+\frac{s_{b,j}}{z_{b,j}}+\frac{s_{b,k}}{z_{b,k}}+\sum_{\tilde{b}\in B\setminus\{b\}}\frac{s_{b,\tilde{b}}}{z_{b,\tilde{b}}}.
\end{align}\end{subequations}
Momentum conservation dictates that
\begin{equation}
    p_k = -p_i -p_j - \sum_{a\in A} p_a -\sum_{b\in B} p_b,
\end{equation}
so we define\begin{subequations}
\begin{align}
    p_\kappa &\equiv p_k+\sum_{b\in B} p_b, \nonumber \\
    & = -p_i -p_j - \sum_{a\in A} p_a,\\
    p_\kappa^\prime & \equiv p_k+\sum_{a\in a} p_a,\nonumber \\
    & = -p_i -p_j - \sum_{b\in B} p_b,
\end{align}\end{subequations}
such that $p_\kappa$ and $p_\kappa^\prime$ are off the mass-shell.
The split scattering equations \eqref{eq:SEsplitting-on} can then be rewritten as\footnote{We emphasize that we define $s_{a,\kappa} = 2\,p_a\cdot p_\kappa$ even for off-shell momenta, implying that in general $s_{a,\kappa}\neq (p_a+p_\kappa)^2$.}
\begin{subequations} \label{eq:SEsplitting-off}
\begin{align}
S_{a}&=\frac{s_{a,i}}{z_{a,i}}+\frac{s_{a,j}}{z_{a,j}}+\frac{s_{a,\kappa}}{z_{a,\kappa}}+\sum_{\tilde{a}\in A\setminus\{a\}}\frac{s_{a,\tilde{a}}}{z_{a,\tilde{a}}},\\
S_{b}&=\frac{s_{b,i}}{z_{b,i}}+\frac{s_{b,j}}{z_{b,j}}+\frac{s_{b,\kappa^\prime}}{z_{b,\kappa^\prime}}+\sum_{\tilde{b}\in B\setminus\{b\}}\frac{s_{b,\tilde{b}}}{z_{b,\tilde{b}}},
\end{align}\end{subequations}
with $z_\kappa = z_{\kappa^\prime} = z_k$. Regarding the measure \eqref{eq:measure}, it is then recast as
\begin{equation}
    \dd\mu_n = \Delta(i,j,\kappa)^{-1} \Delta(i,j,\kappa^\prime)^{-1} \times \dd\mu_{\mathrm{L}}(i, j, A; \kappa) \times \dd\mu_{\mathrm{R}}(i, j, B; \kappa^\prime),   \label{eq:measuresplit}
\end{equation}
with
\begin{align}
    \dd\mu_{\mathrm{L}}(i, j, A; \kappa) &= \Delta(i,j,\kappa)^2 \prod_{a\in A} dz_a \, \delta(S_a),\\
    \dd\mu_{\mathrm{R}}(i, j, B; \kappa^\prime) &= \Delta(i,j,\kappa^\prime)^2 \prod_{b\in B} dz_b \, \delta(S_b).
\end{align}
Here we implicitly use $S_a$ and $S_b$ in equation \eqref{eq:SEsplitting-off}, and $\kappa$ and $\kappa^\prime$ denote off-shell states. Aside from a pre-factor, the CHY measure \eqref{eq:measuresplit} factorizes into two sectors, $\mathrm{L}$ and $\mathrm{R}$.

\subsection{The splitting of the integrands}

Next, we review the splitting behavior of some integrands.
The Parke-Taylor factor in the canonical ordering is defined as
\begin{equation}
\mathrm{PT}({1,2,\dots,n}) = \frac{1}{z_{1,2} \ldots z_{n-1,n} z_{n,1}}.
\end{equation}
Since we are working with $k=n$, it can be conveniently  rewritten  as
\begin{equation}
\mathrm{PT}({1,2,\dots,n})  = \frac{1}{(z_{i,i+1}\ldots z_{j-1,j})}  \times \frac{1}{(z_{1,2}\ldots z_{i-1,i}) (z_{j,j+1}\ldots z_{n-1,n}z_{n,1})}
\end{equation}
Now, if we multiply the first factor on the right-hand side by $1=(z_{j,\kappa} z_{\kappa,i})/(z_{j,\kappa} z_{\kappa,i})$ and the second factor by $1=z_{i,j}/z_{i,j}$, we simply get
\begin{equation}\label{eq:splitPT}
    \mathrm{PT}({1,2,\dots,n}) = \Delta(i,j,\kappa) \underbrace{\mathrm{PT}(i,\ldots,j,\kappa)}_{L}\underbrace{\mathrm{PT}(1,\ldots,i,j,\ldots,\kappa^{\prime})}_{R},
\end{equation}
in which the split behavior is evident.

The CHY representation of Yang--Mills and gravity amplitudes requires the introduction of the $2n\times 2n$  matrix 
\begin{align}
\boldsymbol{\Psi}_n=&{}\left(\begin{array}{cc}
\mathbf{A}_n&-\mathbf{C}_n^\mathrm{T}\\
\mathbf{C}_n&\mathbf{B}_n
\end{array}
\right),\label{MatrixM_n}
\end{align}
\noindent where the  different $n\times n$ sub-matrices are defined as \cite{Cachazo:2013hca,Cachazo:2013iea}
\begin{subequations} \label{eq:SubMatricesM_n}
\begin{align}
\mathbf{A}_{n}^{cd}&=\begin{cases}
\frac{p_{c}\cdot p_{d}}{z_{c,d}}, & \textrm{for }c\neq d,\\
0, & \textrm{for }c=d,
\end{cases}\\\mathbf{B}_{n}^{cd}&=\begin{cases}
\frac{\epsilon_{c}\cdot\epsilon_{d}}{z_{c,d}}, & \textrm{for }c\neq d,\\
0, & \textrm{for }c=d,
\end{cases}\\\mathbf{C}_{n}^{cd}&=\begin{cases}
\frac{\epsilon_{c}\cdot p_{d}}{z_{c,d}}, & \textrm{for }c\neq d,\\
-\sum\limits_{e\neq d}\frac{\epsilon_{d}\cdot p_{e}}{z_{d,e}}, & \textrm{for }c=d,
\end{cases} \label{eq:Cmatrix}
\end{align}\end{subequations}
with  $\epsilon_c$ denoting the $n$ polarization vectors. The half-integrand entering the CHY formula is the so-called   reduced Pfaffian of $\boldsymbol{\Psi}_n$,
\be
\mathrm{Pf}^\prime \boldsymbol{\Psi}_n \equiv \frac{(-1)^{p+q}}{z_{p,q}} \mathrm{Pf} \boldsymbol{\Psi}_n^{[p,q]}, \label{eq:reducedPf}
\ee
where the matrix $\boldsymbol{\Psi}_n^{[p,q]}$ is obtained from $\boldsymbol{\Psi}_n$ by removing the $p$-th and $q$-th rows and columns. The value of $\mathrm{Pf}^\prime \boldsymbol{\Psi}_n$ does not depend on the choice of $p$ and $q$. 

In order to observe the splitting behavior of \eqref{eq:reducedPf}, we need to introduce additional conditions involving the polarization vectors  \cite{Cao:2024gln,Cao:2024qpp}. The first set of conditions is given by
\begin{equation}
    \epsilon_a\cdot p_{b}= p_a\cdot \epsilon_{b} =\epsilon_a\cdot \epsilon_{b}=  0 \qquad \forall a\in A, b\in B.
    \label{eq:extra1}
\end{equation}
They are aligned with the overall intuition of the splitting of the measure, cf. equation \eqref{eq:main}, and also compatible with the residual gauge transformation of the polarizations, $\delta \epsilon_a \propto p_a$ and $\delta \epsilon_b \propto p_b$. The second set of conditions is given by
\begin{equation}
    \epsilon_a \cdot \epsilon_{I} =p_a\cdot \epsilon_{I}=  0 \qquad \forall a\in A, I \in \mathcal{I} = \{i,j,k\}.
    \label{eq:extra2}
\end{equation}
These conditions are not compatible with $\delta \epsilon_I \propto p_I$, although this is a gauge invariance of the reduced Pfaffian \eqref{eq:reducedPf}. We will come back to this point later.

Under the conditions \eqref{eq:extra1} and \eqref{eq:extra2}, the reduced Pfaffian splits as
\begin{equation}
\mathrm{Pf}^\prime \boldsymbol{\Psi}_n \to  \underbrace{\mathrm{Pf}\boldsymbol{\Psi}_{\{A\}}}_{L} \times \underbrace{\mathrm{Pf}^{\prime}\boldsymbol{\Psi}_{\{i,j,B,\kappa^{\prime}\}}}_{R},
\label{eq:splitPf}
\end{equation}
where $\boldsymbol{\Psi}_{\{i_1,\cdots,i_m\}}$ denotes the $2m\times 2m$ sub-matrix of $\boldsymbol{\Psi}_n$ with only rows and columns in $\{i_1,  \ldots , i_m\}$ and $\{i_{1 + n}, , \ldots , i_{m + n}\}$ remaining. Note that only the second term on the right hand side is a reduced Pfaffian.

\subsection{The splitting of the amplitudes}

As a concrete example of the resulting splitting behavior in \eqref{eq:CHYamplitude}, we will analyze the  partial amplitude of Yang--Mills with $n$ external gluons. In the canonical ordering, it is given by
\begin{equation}
{\cal{A}}_n^{\mathrm{YM}}(1,2,\dots,n) = \int_{{\mathbb{C}}^n} \dd\mu_n \,\mathrm{PT}(1,2,\dots,n)\,\mathrm{Pf}^\prime \boldsymbol{\Psi}_n\label{eq:YM-CHY}\,.
\end{equation}
The  splitting conditions \eqref{eq:main}, \eqref{eq:extra1}, and \eqref{eq:extra2} lead to the splitting of the measure, cf. equation \eqref{eq:measuresplit}, the Parke--Taylor factor, cf. equation \eqref{eq:splitPT}, and the reduced Pfaffian, cf. equation \eqref{eq:splitPf}, such that the amplitude \eqref{eq:YM-CHY} is recast as
\begin{equation}\label{eq:YMsplit}
    {\cal{A}}_n^{\mathrm{YM}}(1,2,\dots,n) \to {\cal{J}}^{\mathrm{YM}+\phi^3}(i^\phi,A,j^\phi,\kappa^\phi) \times {\cal{J}}_{\mu}^{\mathrm{YM}}(1,\dots,i,j,\ldots,\kappa^\prime)\epsilon_n^\mu,
\end{equation}
with
\begin{equation}\label{eq:YMscalarcurrent}
    {\cal J}^{\mathrm{YM}+\phi^{3}}(i^\phi,A,j^\phi,\kappa^\phi)=\int\mathrm{d}\mu_{\mathrm{L}}(i,j,A;\kappa)\mathrm{PT}(i,j,\kappa)\mathrm{PT}(i,\ldots,j,\kappa)\mathrm{Pf}\boldsymbol{\Psi}_{\{A\}},
\end{equation}
and
\begin{equation}\label{eq:YMcurrent}
    {\cal J}_{\mu}^{\mathrm{YM}}(1,\dots,i,j,\ldots,\kappa^\prime)\epsilon_n^\mu=\int\mathrm{d}\mu_{\mathrm{R}}(i,j,B;\kappa^{\prime})\mathrm{PT}(1,\dots,i,j,\ldots,\kappa^\prime)\mathrm{Pf}^{\prime}\boldsymbol{\Psi}_{\{j,B,\kappa^{\prime},i\}}.
\end{equation}
${\cal{J}}^{\mathrm{YM}+\phi^3}$ in \eqref{eq:YMscalarcurrent} denotes the scalar current with $|A|$ gluons and  two scalars ($i,j$), with  off-shell leg $\kappa$, while ${\cal{J}}_\mu^{\mathrm{YM}}$ denotes a pure gluon current with off-shell leg $\kappa^\prime$. 

Notice that in the latter, the residual gauge invariance of the legs $i$, $j$, and $k=n$ is lost. This  can be illustrated with a simple example. Consider the Yang-Mills amplitude $ \mathcal{A}_4^{\mathrm{YM}}(1,2,3,4)$, and the sets $A=\{2\}$  and $B=\emptyset$. 
In this case, the splitting conditions reduce to,
\begin{equation}
\epsilon_2 \cdot \epsilon_I = p_2 \cdot \epsilon_I = 0,
\qquad I=\{1,3,4\}.
\end{equation}
One then obtains
\begin{align}\label{eq:YM4point}
\mathcal{A}_4^{\mathrm{YM}}(1,2,3,4)  &\to {\cal{J}}^{\mathrm{YM}+\phi^3}(1^\phi,2,3^\phi,\kappa^\phi) \times {\cal{J}}_{\mu}^{\mathrm{YM}} (1,3,\kappa^\prime)\epsilon_4^\mu \nonumber\\
& = \Big( \frac{ p_{1} {\cdot} \epsilon_2}{s_{1,2}} - \frac{ p_3 {\cdot} \epsilon_2 }{s_{2,3}} \Big)\times \left( (\epsilon_1\cdot \epsilon_3) (p_1\cdot \epsilon_4) + (\epsilon_3\cdot \epsilon_4) (p_3\cdot \epsilon_1)+(\epsilon_4\cdot \epsilon_1) (p_{24}\cdot \epsilon_3)\right),
\end{align}
with $p_{24}\equiv p_2+p_4$. From this expression, it is clear that the amplitude after splitting does not vanish when one replaces $\epsilon_I \to p_I$ for $I=\{1,3,4\}$. In contrast, setting $\epsilon_2 \to p_2$ yields zero. This explicitly shows that the residual gauge invariance is preserved only for leg $2$, while it is lost for the remaining legs in this configuration.

This was already expected because of the splitting conditions in \eqref{eq:extra2}. However, when we put the off-shell object \eqref{eq:YMcurrent} on the mass-shell, i.e. by imposing $p_\kappa^\prime = p_n$, and momentum conservation, it is again identified with a tree level amplitude of the Yang-Mills theory, with residual gauge invariance restored.

\section{Splitting in higher-derivative theories}\label{sec:HDtheories}

In reference \cite{Azevedo:2017lkz}, it was shown that the tree level scattering amplitudes for the plane-wave states of the $(DF)^2$ theory, i.e. gluon-like excitations, admit a CHY representation. In the canonical ordering, it is given by
\begin{equation}\label{eq:CHY-DF2}
{\cal{A}}_n^{(DF)^2}(1,2,\dots,n) = \int_{{\mathbb{C}}^n} \dd\mu_n \,\mathrm{PT}(1,2,\dots,n)\,W_{\{1,2,\dots,n\}},
\end{equation}
%
where $W_{\{1,2,\dots,n\}}$ denotes a product of the Lam-Yao cycles~\cite{PhysRevD.93.105008}, usually denoted by $W_{11\cdots1}$ in the literature. These cycles are common building blocks of CHY amplitudes for theories with higher-dimensional operators. In our case, $W_{\{1,2,\dots,n\}}$ can be written in terms of the diagonal elements of the $\mathbf{C}$ matrix \eqref{eq:Cmatrix},
\begin{equation}\label{eq:Wn}
W_{\{1,2,\dots,n\}} = \prod_{c=1}^n \left(-\sum\limits_{d\neq c}\frac{\epsilon_{c}\cdot p_{d}}{z_{c,d}}\right).
\end{equation}
More generally, we will work with the following operator,
\begin{equation}
    W_{\{i_1,\ldots,i_m\}} \equiv \prod_{r=1}^m \left(- \sum_{\substack{q=1\\(q\neq r)}}^{m} \frac{\epsilon_{i_r}\cdot p_{i_q}}{z_{i_r,i_q}}\right).
\end{equation}
From this ingredient, it is clear that the scattering amplitudes \eqref{eq:CHY-DF2} do not involve contracted polarizations, $\epsilon_c \cdot \epsilon_d$. This is  a notable property of the $(DF)^2$ theory~\cite{Johansson:2017srf}.
%

\subsection{Splitting in the  $(DF)^2$ theory}

Regarding the splitting discussed in section \ref{sec:review}, after imposing the conditions \eqref{eq:extra1} and \eqref{eq:extra2}, it is trivial to show that
\begin{equation}
W_{\{1,2,\dots,n\}} \to \underbrace{\prod_{a\in A}\left(-\sum_{\substack{\tilde{a}\in A\cup\{i,j,\kappa\}\\
(\tilde{a}\neq a)
}
}\frac{\epsilon_{a}\cdot p_{\tilde{a}}}{z_{a,\tilde{a}}}\right)}_{L}\underbrace{W_{B\cup\{i,j,\kappa^{\prime}\}}}_{R},
\label{eq:splitW}
\end{equation}
which is analogous to equation \eqref{eq:splitPf}.
The  splitting \eqref{eq:splitW}, together with equations \eqref{eq:measuresplit} and \eqref{eq:splitPT}, implies that the $(DF)^2$ amplitude \eqref{eq:CHY-DF2} with $n$ external gluons splits as
\begin{equation}\label{eq:DF2-splitting_}
{\cal{A}}_n^{(DF)^2}(1,2,\dots,n) \to {\cal{J}}^{(DF)^2+\phi^3}(i^\phi,A,j^\phi,\kappa^\phi) \times {\cal{J}}_{\mu}^{(DF)^2}(1,\dots,i,j,\ldots,\kappa^\prime)\epsilon_n^\mu,
\end{equation}
where the currents ${\cal{J}}^{(DF)^2+\phi^3}$ and ${\cal{J}}_{\mu}^{(DF)^2}$ are the $DF^2$ analogous of the Yang-Mills currents \eqref{eq:YMscalarcurrent} and \eqref{eq:YMcurrent}, respectively.

The scattering amplitudes in these $(DF)^2$ theories can be derived from a Lagrangian introduced in \cite{Johansson:2017srf}, given by
\begin{multline}\label{eq:DF2phi3}
\mathcal{L}_{(DF)^2+\phi^3} =  \frac{1}{2}(D_{\mu} F^{\mu \nu\,I})^2  + \frac{g}{3} \, F^3+ \frac{1}{2}(D_{\mu} \varphi^{\alpha})^2  + \frac{g}{2}  \,  C^{\alpha IJ}  \varphi^{ \alpha}   F_{\mu \nu}^I F^{\mu \nu\,J } +  \frac{g}{3!}  \, d^{\alpha \beta \gamma}   \varphi^{ \alpha}  \varphi^{ \beta} \varphi^{ \gamma} \\ + \frac{1}{2} (D_{\mu} \phi^{I\widehat{I}})^2 + \frac{g\lambda}{3!} f^{IJK} \hat f^{\widehat{I}\widehat{J}\widehat{K}} \phi^{I\widehat{I}} \phi^{J\widehat{J}} \phi^{K\widehat{K}} + \frac{g}{2} \varphi^\alpha  \phi^{I\widehat{K}} \phi^{J\widehat{K}} C^{\alpha IJ}.  
\end{multline}
Observe that the massless scalars $\phi^{I\widehat{I}}$ are charged under two groups: while the gauge group (adjoint) indices $I,J,K,\ldots$ are shared with the gluons, the second group (global) with indices $\widehat{I},\widehat{J},\widehat{K},\ldots$ is exclusive to $\phi^{I\widehat{I}}$.
The field strength and covariant derivatives are defined as\begin{subequations}
\begin{align}
F_{\mu \nu}^I &= \partial_{\mu} A_{\nu}^I-\partial_{\nu} A_{\mu}^I + g  f^{IJK} A_{\mu}^J A_{\nu}^K,  \\
F^3 &=  f^{IJK} F_{\mu}^{I\, \nu}F_{\nu}^{J \, \lambda} F_{\lambda }^{K\, \mu},\\
D_{\mu} \varphi^\alpha & = \partial_{\mu} \varphi^\alpha -  i g  (T_{R}^{I})^{\alpha \beta} A_{\mu}^I \varphi^\beta, \\
D_{\rho} F_{\mu \nu}^I &= \partial_{\rho} F_{\mu \nu}^I +  g  f^{IJK} A_{\rho}^J F_{\mu \nu}^K,\\
D_{\mu} \phi^{I\widehat{I}} &= \partial_{\mu} \phi^{I\widehat{I}} +  g  f^{IJK} A_{\mu}^J \phi^{K\widehat{I}}.
\end{align}\end{subequations}
The scalar $\varphi^\alpha$ transforms in a real representation $R$ with indices $\alpha,\beta,\gamma,\ldots$, and $g$ is the coupling constant.
The Clebsh--Gordan coefficients $C^{\alpha IJ}$ and $d^{\alpha \beta \gamma}$ are implicitly defined through the two relations
\begin{align}
C^{\alpha IJ}C^{\alpha KL} & = f^{IKM}f^{MLJ}+ f^{ILM}f^{MKJ}, \\
C^{\alpha IJ}d^{\alpha \beta \gamma} &= (T_{R}^I)^{\beta \alpha} (T_{R}^J)^{\alpha \gamma}+ C^{\beta IK} C^{\gamma KJ} + (I \leftrightarrow J). 
\end{align}
Interestingly, these relations imply that pure gluon amplitudes can be color-ordered.

A simple instance of the  splitting \eqref{eq:DF2-splitting_} can already be seen at $n=4$. Indeed, using equation \eqref{eq:CHY-DF2}, we obtain
\begin{multline}
{\cal{A}}_4^{(DF)^2}(1,2,3,4) =  -\frac{s_{1,2}^{2}s_{2,3}^{2}}{s_{1,3}}\\ \times \left(\frac{p_{2}\cdot\epsilon_{1}}{s_{1,2}}-\frac{p_{4}\cdot\epsilon_{1}}{s_{2,3}}\right)\left(\frac{p_{1}\cdot\epsilon_{2}}{s_{1,2}}-\frac{p_{3}\cdot\epsilon_{2}}{s_{2,3}}\right)\left(\frac{p_{4}\cdot\epsilon_{3}}{s_{1,2}}-\frac{p_{2}\cdot\epsilon_{3}}{s_{2,3}}\right)\left(\frac{p_{3}\cdot\epsilon_{4}}{s_{1,2}}-\frac{p_{1}\cdot\epsilon_{4}}{s_{2,3}}\right).
\label{eq:4pt-DF2}
\end{multline}
Now, choosing the reference particles to be $i=1$, $j=3$ and $k=4$,   we are left with the sets $A=\{2\}$ and $B=\emptyset$. In this case, the splitting conditions are simply $p_2\cdot\epsilon_I  = 0$, with $I \in \{1,3,4\}$. The amplitude \eqref{eq:4pt-DF2} then becomes
\begin{equation}
{\cal{A}}_4^{(DF)^2}(1,2,3,4) \to\left(\frac{p_{3}\cdot\epsilon_{2}}{s_{2,3}}-\frac{p_{1}\cdot\epsilon_{2}}{s_{1,2}}\right)(p_{3}\cdot\epsilon_{1})(p_{1}\cdot\epsilon_{3})(p_{3}\cdot\epsilon_{4}),
\label{eq:4ptDF2split}
\end{equation}
where we have used momentum conservation and the on-shell conditions $p_q^2= p_q\cdot\epsilon_q=0$. It is easy to check that
\begin{equation}
    {\cal J}^{(DF)^{2}+\phi^{3}}(1^{\phi},2,3^{\phi},\kappa^{\phi})=\left(\frac{p_{1}\cdot\epsilon_{2}}{s_{1,2}}-\frac{p_{3}\cdot\epsilon_{2}}{s_{2,3}}\right),
\end{equation}and
\begin{equation}
    {\cal J}_{\mu}^{(DF)^{2}}(1,3,\kappa^{\prime})=\frac{1}{2}(p_{3}\cdot\epsilon_{1})(p_{1}\cdot\epsilon_{3})(p_{1\mu}-p_{3\mu}).
\end{equation}
Therefore, the splitting \eqref{eq:4ptDF2split} agrees with the general equation \eqref{eq:DF2-splitting}.

In \cite{Cao:2024qpp}, an analogous splitting was considered for $\mathcal{A}_4^{\mathrm{YM}}(1,2,3,4)$, with
\begin{equation}
\Big( \frac{ p_{1} {\cdot} \epsilon_2}{s_{1,2}} - \frac{ p_3 {\cdot} \epsilon_2 }{s_{2,3}} \Big) =  \mathcal{J}^{\mathrm{YM}+\phi^3}(1^\phi,2,3^\phi,\kappa^\phi).
\end{equation}
Indeed, $\mathcal{J}^{\mathrm{YM}+\phi^3}(1^\phi,2,3^\phi,\kappa^\phi)=\mathcal{J}^{(DF)^2+\phi^3}(1^\phi,2,3^\phi,\kappa^\phi)$, since both expressions come from the Berends--Giele currents derived from the terms $ \frac{1}{2} (D_{\mu} \phi^{I\widehat{I}})^2 + (\phi^{I\widehat{I}})^3$ in the respective Lagrangians. Accordingly, note that $\mathrm{Pf}\boldsymbol{\Psi}_{\{2\}}= \mathbf{C}_{\{2\}}^{1,1} $.

\subsection{Gravity theories}

In \cite{Cao:2024gln,Cao:2024qpp}, the splitting  of Einstein gravity amplitudes was also investigated. In the CHY representation, they are cast as
\begin{equation}
\mathcal{M}_n^{\mathrm{GR}} = \int_{{\mathbb{C}}^n} \dd\mu_n \,\mathrm{Pf}^\prime \boldsymbol{\Psi}_n\,\mathrm{Pf}^\prime \boldsymbol{\Psi}_n\label{GR},
\end{equation}
where the graviton polarization is written as $h_n^{\mu \nu} = \epsilon^{\mu}_n \epsilon^{\nu}_n$. The presence of two copies of the Pfaffian now allows for two distinct choices, leading to different splittings. One could  split the Pfaffian by imposing either (1) the conditions in equations \eqref{eq:extra1} and  \eqref{eq:extra2}, or (2) the conditions in equations \eqref{eq:extra1} and
\begin{equation}
    \epsilon_b \cdot \epsilon_{I} =p_b\cdot \epsilon_{I}=  0 \qquad \forall b\in B, I \in \mathcal{I}.
    \label{eq:extra3}
\end{equation}
Under the latter, the Pfaffian splits as
\begin{equation}
\mathrm{Pf}^\prime \boldsymbol{\Psi}_n \to   \underbrace{\mathrm{Pf}^{\prime}\boldsymbol{\Psi}_{\{i,A,j,\kappa^{\prime}\}}}_{L}\times \underbrace{\mathrm{PT}(i,j,\kappa)\mathrm{Pf}\boldsymbol{\Psi}_{\{B\}}}_{R},
\label{eq:splitPfalt}
\end{equation}

For Yang-Mills amplitudes, the different splitting choices simply reverse the roles of the sets $A$ and $B$. For Einstein gravity amplitudes, on the other hand, the two options lead to the following splittings,
\begin{equation}
\mathcal{M}_n^{\mathrm{GR}} \to \mathcal{J}^{\mathrm{GR}+\phi^3}(i^\phi,A,j^\phi,\kappa^\phi) \times \mathcal{J}_{\mu\nu}^{\mathrm{GR}}(1,\dots,i,j,\ldots,\kappa^\prime)\epsilon_n^\mu\epsilon_n^\nu
\label{eq:GRsplit}
\end{equation}
or
\begin{equation}
\mathcal{M}_n^{\mathrm{GR}} \to \mathcal{J}_{\mu}^{\mathrm{EYM}}(i^g,A,j^g,\kappa^g)\epsilon_n^\mu \times \mathcal{J}_{\nu}^{\mathrm{EYM}}(1,\dots,i^g,j^g,\ldots,n-1,\kappa^{\prime g})\epsilon_n^\nu,
\label{eq:GRsplitalt}
\end{equation}
where EYM stands for Einstein-Yang-Mills and the superscript $g$ indicates a gluonic leg. 

\subsubsection{$R^2$ gravity}

Graviton amplitudes for the curvature-squared gravity coming from heterotic ambitwistor strings (which reduce to conformal gravity amplitudes in four dimensions) admit the following CHY representation \cite{Azevedo:2017lkz},
\begin{equation}
\mathcal{M}_n^{R^2} = \int_{{\mathbb{C}}^n} \dd\mu_n \,\mathrm{Pf}^\prime \boldsymbol{\Psi}_nW_{\{1,2,\dots,n\}}\label{R^2}.
\end{equation}

Given the splittings found for the reduced Pfaffian, cf. equation~\eqref{eq:splitPf}, and for $W_{\{1,2,\dots,n\}}$, cf. equation~\eqref{eq:splitW}, there are two possible splittings for $\mathcal{M}_n^{R^2}$. By imposing the conditions \eqref{eq:main}, \eqref{eq:extra1} and \eqref{eq:extra2}, we obtain
\begin{equation}
\mathcal{M}_n^{R^2} \to \mathcal{J}^{R^2+\phi^3}(i^\phi,A,j^\phi,\kappa^\phi) \times \mathcal{J}_{\mu\nu}^{R^2}(1,\dots,i,j,\ldots,\kappa^\prime)\epsilon_n^\mu\epsilon_n^\nu,
\label{eq:R2split}
\end{equation}
which is analogous to \eqref{eq:GRsplit}. On the other hand, if we split $W_{\{1,\ldots,n\}}$ as in \eqref{eq:splitW} while choosing the alternative splitting \eqref{eq:splitPfalt} for the Pfaffian, we get
\begin{equation}
\mathcal{M}_n^{R^2} \to \mathcal{J}_{\mu}^{R^2+\mathrm{YM}}(i^g,A,j^g,\kappa^g)\epsilon_n^\mu \times \mathcal{J}_{\nu}^{R^2+(DF)^2}(1,\dots,i^g,j^g,\ldots,n-1,\kappa^{\prime g})\epsilon_n^\nu\,.
\label{eq:R2splitalt}
\end{equation}
The mixed currents above can be obtained from their respective CHY representations \cite{Azevedo:2018dgo}. 

\subsubsection{$R^3$ gravity}

Finally, we can analyze the splitting of the graviton amplitude in the higher derivative gravity theory described by bosonic ambitwistor strings \cite{Azevedo:2017lkz}, which we will call $R^3$. The corresponding tree level amplitudes can be computed using the perturbiner method in the $\alpha^\prime \to 0$ limit of the equations of motion recently derived in \cite{Carabine:2023yxv}. Their CHY representation is given by
\begin{equation}
\mathcal{M}_n^{R^3} = \int_{{\mathbb{C}}^n} \dd\mu_n \,W_{\{1,2,\dots,n\}}W_{\{1,2,\dots,n\}}\label{eq:CHY-R^3}.
\end{equation}

Regarding the possible splittings, this case is completely analogous to the graviton amplitudes discussed so far. By imposing the splitting conditions \eqref{eq:extra2} in both $W_{\{1,2,\dots,n\}}$, we obtain
\begin{equation}
\mathcal{M}_n^{R^3} \to \mathcal{J}^{R^3+\phi^3}(i^\phi,A,j^\phi,\kappa^\phi) \times \mathcal{J}_{\mu\nu}^{R^3}(1,\dots,i,j,\ldots,\kappa^\prime)\epsilon_n^\mu\epsilon_n^\nu.
\label{eq:R3split}
\end{equation}
Alternatively, by imposing \eqref{eq:extra2} on one $W_{\{1,\ldots,n\}}$ and \eqref{eq:extra3} on the other leads to
\begin{equation}
\mathcal{M}_n^{R^3} \to \mathcal{J}_{\mu}^{R^3+(DF)^2}(i^g,A,j^g,\kappa^g)\epsilon_n^\mu \times \mathcal{J}_{\nu}^{R^3+(DF)^2}(1,\dots,i^g,j^g,\ldots,n-1,\kappa^{\prime g})\epsilon_n^\nu,
\label{eq:R3splitalt}
\end{equation}
where the superscript  $g$ denotes gluon states from the  $(DF)^2$ theory.


\section{On-Shell representations of the split amplitudes} \label{sec:on-shellrep}

The splitting of a given amplitude is cast in terms of amputated currents, which are usually gauge dependent objects. Indeed, they possess a nontrivial structure due to the off-shell component. Their computation is a bit more subtle within the CHY framework. It would be more interesting to recast the splitting in terms of physical objects, i.e. lower-point amplitudes. In this section, we demonstrate how to do this with an appropriate choice of kinematic data.

\subsection{An example}\label{sec:example.On-shell}

Let us consider the five-point bi-adjoint (BA) amplitude with $(i,j,k) = (1,3,5)$, where $A = \{2\}$ and $B = \{4\}$. A direct computation leads to 
\begin{align}
\mathcal{A}_5^{\phi^3}(1,2,3,4,5)&=\frac{1}{s_{1,2}s_{3,4}}+\frac{1}{s_{1,2}s_{4,5}}+\frac{1}{s_{2,3}s_{4,5}}+\frac{1}{s_{2,3}s_{1,5}}+\frac{1}{s_{1,5}s_{3,4}}\nonumber\\
& \to \left(\frac{1}{s_{1,2}}+\frac{1}{s_{2,3}}\right) \left( \frac{1}{s_{3,4}}+\frac{1}{s_{4,5}}  \right). \label{eq:CHY5pts}		
\end{align}
The transition to the second line is simply the imposition of the splitting condition \eqref{eq:main}, that is, $s_{2,4}=0$. As shown in \cite{Cao:2024gln,Cao:2024qpp}, the 2-split representation for this configuration is given by
\begin{align}
\mathcal{A}_5^{\phi^3}(1,2,3,4,5) \to \mathcal{J}^{\phi^3}(1,2,3,\kappa)\, \mathcal{J}^{\phi^3}(1,3,4,\kappa^\prime) \label{eq:2splitBA5},		
\end{align} 
with off-shell legs $p_\kappa=-p_1-p_2-p_3=p_4+p_5$ and $p_{\kappa^\prime}=-p_1-p_3-p_4=p_2+p_5$.
Notice that the amputated current $\mathcal{J}^{\phi^3}(1,2,3,\kappa)$ is directly related to the bi-adjoint Berends--Giele (BG) current \cite{Mafra:2016ltu},
\begin{align}
\mathcal{J}_{\rm CHY}^{\phi^3}(1,2,3,\kappa) &= s_{1,2,3}\mathcal{J}_{\rm BG}^{\phi^3}(1,2,3,\kappa), \nonumber\\
&= \frac{1}{s_{1,2}}+\frac{1}{s_{2,3}}\label{eq:CHY5pts1},
\end{align}
in which the puncture $z_2$ was integrated. In contrast, the current $\mathcal{J}^{\phi^3}(1,3,4,\kappa^\prime)$ is evaluated with an integration over $z_4$. This difference becomes relevant in the comparison with the respective BG current. Indeed,
\begin{align}
\mathcal{J}_{{\rm CHY}}^{\phi^3}(1,3,4,\kappa^\prime) &
= \frac{1}{s_{3,4}}+\frac{1}{s_{4,\kappa^\prime}}
\nonumber\\
&=\frac{1}{s_{3,4}}+\frac{1}{s_{4,5}}, \label{eq:CHY5pts2}\\
s_{1,3,4} \mathcal{J}_{{\rm BG}}^{\phi^3}(1,3,4,\kappa^\prime) &= \frac{1}{s_{3,4}}+\frac{1}{s_{1,3}}  \nonumber\\
&=\frac{1}{s_{3,4}}+\frac{1}{s_{2,5}+s_{4,5}} , \label{eq:BG5pts}
\end{align}
where we have used the condition $s_{2,4}=0$. 
As expected, the product of equations \eqref{eq:CHY5pts1} and \eqref{eq:CHY5pts2} correctly reproduces the result in \eqref{eq:CHY5pts}. However, there is a clear discrepancy between equations \eqref{eq:CHY5pts2} and \eqref{eq:BG5pts}, namely, $\mathcal{J}_{{\rm CHY}}^{\phi^3}(1,3,4,\kappa^\prime) \neq s_{1,3,4}\,{\cal J}_{\rm BG}^{\phi^3}(1,3,4,\kappa^\prime)$.

The mismatch between the CHY and BG currents vanishes when $\kappa^\prime$ is put on-shell, with $p_{\kappa}^{_{'}2} = s_{2,5} = 0$. Related ideas were further discussed by one of the authors in \cite{Gomez:2025tqx}, where it was demonstrated that amputated CHY currents for bi-adjoint and Yang--Mills theories can be derived directly from amplitudes. The same procedure can be straightforwardly extended to the other theories analyzed here (see appendix \ref{IMI}).

In the following subsection, we adopt the method proposed in \cite{Naculich:2014naa,Naculich:2015zha} which enables the construction of CHY currents via on-shell methods.

\subsection{Kinematic shifting}\label{sec.kinematic}

Let us consider CHY amputated currents with up to three off-shell external legs, with  
$p_I^2\neq 0$, $I \in \{i,j,\kappa\}$, and $(n-3)$ on-shell legs. The ${\rm SL}(2,\mathbb{C})$ invariant scattering equations $S_c=0$ are modified as \cite{Naculich:2014naa,Naculich:2015zha}
\begin{equation}\label{eq:MSE}
S_c=\sum_{d\neq c}^n\frac{s_{c,d} + \Delta_{c,d} }{z_{c,d}},
\end{equation}
where the only non-vanishing entries of $\Delta_{c,d}$ are
\begin{subequations} \label{eq:DeltaMSE}
\begin{align} 
\Delta_{i,j}=\Delta_{j,i} & =p_i^2+p_j^2-p_\kappa^2,\\
\Delta_{i,\kappa}=\Delta_{\kappa,i}& =p_i^2+p_\kappa^2-p_j^2, \\ \Delta_{j,\kappa}=\Delta_{\kappa,j} &=p_j^2+p_\kappa^2-p_i^2.
\end{align}
\end{subequations}

Using the ${\rm SL}(2,\mathbb{C})$ symmetry, we can eliminate the three scattering equations associated with the off-shell legs. As a result, the gauge fixed  measure has the same structure as in the case where all external particles are massless. 


For the BA theory, the half-integrands in the CHY representation do not depend on momentum variables. Therefore, we could simply compute amplitudes using the usual scattering equations and shift the kinematic data. In the analysis of the amplitude splitting, the resulting currents have only one off-shell leg, $\kappa$ or $\kappa^\prime$. So we are going to define the shifting operation
\begin{equation}\label{eq:shiftDelta}
    \Sigma(\kappa) \equiv \{s_{c,d} \mapsto s_{c,d} + \Delta_{c,d}\},
\end{equation}
with $\Delta_{c,d}$ given in equation \eqref{eq:DeltaMSE}, now with $p_i^2=p_j^2=0$.

For example, let us consider the 2-split case presented in equation \eqref{eq:CHY5pts}. The identification of the CHY current with the BG current in equation \eqref{eq:CHY5pts1} remains,
\begin{align}
\mathcal{A}_4^{\phi^3}(1,2,3,\kappa)&= \lim_{s_{1,2,3}\to 0}  s_{1,2,3} \mathcal{J}_{\rm BG}^{\phi^3}(1,2,3,\kappa),\nonumber\\
\mathcal{J}^{\phi^3}(1,2,3,\kappa) & =
\left.\mathcal{A}_4^\phi(1,2,3,\kappa)\right|_{\Sigma(\kappa)} \nonumber \\
& = \frac{1}{s_{1,2}}+\frac{1}{s_{2,3}}.
\end{align}
On the other hand, it is easy to show that the (CHY) current  $\mathcal{J}^{\phi^3}(1,3,4,\kappa^\prime)$ can be derived from the respective amplitude when applying the kinematic shift,
\begin{align}
\mathcal{A}_4^{\phi^3}(1,3,4,\kappa^\prime) &= \lim_{s_{1,3,4}\to 0} s_{1,3,4} \mathcal{J}_{{\rm BG}}^{\phi^3}(1,3,4,\kappa^\prime), \nonumber \\
\mathcal{J}^{\phi^3}(1,3,4,\kappa^\prime) &=
\left. \mathcal{A}_4^{\phi^3}(1,3,4,\kappa^\prime)\right|_{\Sigma(\kappa^\prime)}\nonumber \\
 &=\frac{1}{s_{3,4}}+\frac{1}{s_{1,3}-p_{\kappa^\prime}^2}\nonumber \\
 & =\frac{1}{s_{3,4}}+\frac{1}{s_{4,\kappa^\prime}}.
\end{align}

More generally, the splitting of the tree level scattering amplitudes for the bi-adjoint scalar theory can be expressed as
\begin{equation}
\mathcal{A}^{\phi^3}(1,2,\dots,n) \to  \left.\mathcal{A}^{\phi^3}(i,A,j,\kappa) \right|_{\Sigma(\kappa)} \times \left.\mathcal{A}^{\phi^3}(1,\ldots,i,j,\ldots,\kappa^\prime) \right|_{\Sigma({\kappa^\prime)}} \label{}.		
\end{equation} 
Now we should   check whether this kinematic shift can be extended to the other half-integrands of the CHY representations, since they do involve kinematic data.

We start by analyzing $W_{\{1,2,\dots,n\}}$, cf. equation \eqref{eq:Wn}, within the amputated currents. Since it involves only $p_c \cdot \epsilon_d$, the off-shell shift represented by $\Delta_{c,d}$ does not modify it. Additionally, the split conditions imply that the polarization vector associated with the off-shell leg, $\epsilon_{\kappa^\prime}$, is transversal. Therefore, for this class of integrands it suffices to evaluate the associated split amplitudes and implement the shift \eqref{eq:shiftDelta}. For example, the split amplitude of equation \eqref{eq:DF2-splitting} is recast as 
\begin{multline}
\mathcal{A}^{(DF)^2}(1,2,\dots,n) \\ \to
\left. \mathcal{A}^{(DF)^2+\phi^3}(i^\phi,A,j^\phi,\kappa^\phi)\right|_{\Sigma(\kappa)} \times \left.\mathcal{A}^{(DF)^2}(1,\ldots,i,j,\ldots,\kappa^\prime)\right|^{\epsilon_{\kappa^\prime}\rightarrow \epsilon_n}_{\Sigma(\kappa^\prime)}.
\end{multline}

The second half-integrand of interest is the reduced Pfaffian, cf. equation \eqref{eq:reducedPf}. The matrix $\boldsymbol{\Psi}_n$ is composed by the sub-matrices $\mathbf{A}_n$, $\mathbf{B}_n$, and $\mathbf{C}_n$. The latter remains well-defined with the shift, being consistent with the splitting conditions and the shift operation \eqref{eq:shiftDelta}. The sub-matrix $\mathbf{B}_n$ involves only the polarization vectors, so it also left untouched. Finally, the sub-matrix $\mathbf{A}_n$ is the only with a similar structure to the scattering equations. The idea then is to simply perform the shift \eqref{eq:shiftDelta} in its entries, such that  
\begin{equation}
    \left. \mathbf{A}_{n}^{cd}\right|_{\Sigma{(\kappa)}}=\begin{cases}
\frac{s_{c,d}+\Delta_{c,d}}{2\, z_{c,d}}, & \textrm{for }c\neq d,\\
0, & \textrm{for }c=d.
\end{cases}
\end{equation}
In analogy with the bi-adjoint scalar and the $(DF)^2$ theory, for this class of half-integrands it  suffices  to compute the amplitudes and then implement the corresponding shift conditions in order to express the off-shell currents. For pure Yang–Mills, the splitting can be written as
\begin{equation}
\mathcal{A}^\mathrm{YM}(1,2,\dots,n) \to
\left. \mathcal{A}^{\mathrm{YM}+\phi^3}(i^\phi,A,j^\phi,\kappa^\phi)\right|_{\Sigma(\kappa)}  \times 
\mathcal{A}^{\mathrm{YM}}(1,\ldots,i,j,\ldots,\kappa^\prime)\Big|^{\epsilon_{\kappa^\prime}\rightarrow \epsilon_n}_{\Sigma(\kappa^\prime)},
\end{equation}
validating the underlying idea of a kinematic shifting.

\section{Transmuting operators}

Originally introduced in \cite{Cheung:2017ems}, transmuting operators are differential operators that map tree-level scattering amplitudes from one theory to those of another. In this section, we introduce a new transmuting operator. Among other properties, it emulates the coupling to bi-adjoint scalars in the amplitudes of the different higher-derivative field theories discussed in section \ref{sec:HDtheories}.

Inspired by the transmuting operators defined in \cite{Cheung:2017ems}, given by
\begin{align}
\mathcal{T}^{\Psi}[i_1,i_2] & =  \mathcal{T}_{i_1i_2},\\
\mathcal{T}^{\Psi}[i_1,i_2,\ldots, i_m] &= \left(\prod_{a=2}^{m-1}  \mathcal{T}_{i_{a-1}i_ai_m}\right)   \mathcal{T}_{i_1i_m} , 	
\end{align}
for $3 \leq m\leq n$, with
\begin{equation}
\mathcal{T}_{ij} = \partial_{\epsilon_i\cdot \epsilon_j}, \qquad
\mathcal{T}_{ijr} \equiv \partial_{p_i\cdot \epsilon_j} - \partial_{\epsilon_j\cdot p_r}, 	
\end{equation}
we define
\begin{equation}
    \mathcal{T}^W[i_1, i_2] =  \mathcal{T}_{i_1i_2i_3} \cdot \mathcal{T}_{i_2i_1i_3},
\end{equation}
where the label $i_3$ works as an auxiliary momentum, and
\begin{equation}
\mathcal{T}^W[i_1,i_2,\ldots, i_m] = \left(\prod_{a=2}^{m-1}  \mathcal{T}_{i_{a-1}i_ai_m}\right) \mathcal{T}_{i_1i_mi_2} \mathcal{T}_{i_mi_1i_2},
\end{equation}
also for $3 \leq m\leq n$.
It is then straightforward to verify the following identities,\begin{subequations}
\begin{align}
\mathcal{T}^{\Psi}[i_1,i_2,\ldots, i_m] \cdot \mathrm{Pf}'\boldsymbol{\Psi}_n  &= \mathrm{PT}(i_1,i_2,\ldots, i_m) \mathrm{Pf}\boldsymbol{\Psi}_{\{ H\}}, \\
\mathcal{T}^{W}[i_1,i_2,\ldots, i_m] \cdot \mathrm{Pf}'\boldsymbol{\Psi}_n&=0,\\
\mathcal{T}^{\Psi}[i_1,i_2,\ldots, i_m] \cdot W_{\{1,2,\ldots,n\}}& =0,
\\
\mathcal{T}^{W}[i_1,i_2,\ldots, i_m] \cdot W_{\{1,2,\ldots,n\}} &= \mathrm{PT}(i_1,i_2,\ldots, i_m)  \prod_{h\in H} {\bf C}_n^{hh},
\end{align}\end{subequations}
where $H=\{1,2,\ldots,n\}\setminus \{ i_1,i_2,\ldots, i_m\} $.
Because of these properties, the action of the  transmuting operators gives rise to several relations between amplitudes in different theories, such as
\begin{align}
\mathcal{A}^{\phi^3}(1,2,\ldots, n)&= \mathcal{T}^W[1,2,\ldots,n] \mathcal{A}^{(DF)^2}(1,2,\ldots,n),\\
\mathcal{A}^{(DF)^2}(1,2,\ldots, n)&= \mathcal{T}^W[1,2,\ldots,n]  \mathcal{M}^{R^3}(1,2,\ldots,n),\\	
\mathcal{A}^{(DF)^2}(1,2,\ldots, n)&= \mathcal{T}^{\Psi}[1,2,\ldots,n]  \mathcal{M}^{R^2}(1,2,\ldots,n),	
\end{align}
and
\begin{equation}
    \mathcal{A}^{(DF)^2+\phi^3}(1^\phi,\ldots,i^\phi,i+1,\ldots, n)= \mathcal{T}^W[1,2\ldots,i] \mathcal{A}^{(DF)^2}(1,2,\ldots,n). 
\end{equation}

Since we have reformulated the splitting operation in terms of sub-amplitudes, we are set to extend the action of the transmuting operators to the split amplitudes. Below we list some examples:
\begin{multline}
    \mathcal{A}^\mathrm{YM}(1,2,\dots,n)\to
\mathcal{T}^\Psi[i,j,\kappa]
 \left.\mathcal{A}^\mathrm{YM}(i,A,j,\kappa)\vphantom{\mathcal{A}^{R^3}}\right|_{\Sigma(\kappa)} \times \left.\mathcal{A}^\mathrm{YM}(1,\ldots,i,j,\ldots, \kappa^\prime)\vphantom{\mathcal{A}^{R^3}}\right|^{\epsilon_{\kappa^\prime}\to \epsilon_n}_{\Sigma(\kappa^\prime)},
\end{multline}

\begin{multline}
    \mathcal{A}^{(DF)^2}(1,2,\dots,n) \to 
\mathcal{T}^W[i,j,\kappa]
\left. \mathcal{A}^{(DF)^2}(i,A,j,\kappa)\right|_{\Sigma(\kappa)} \times  \left.\mathcal{A}^{(DF)^2}(1,\ldots,i,j,\ldots, \kappa^\prime)\right|^{\epsilon_{\kappa^\prime}\to \epsilon_n}_{\Sigma(\kappa^\prime)},
\end{multline}

\begin{multline}
    \mathcal{M}_n^\mathrm{GR} \to
\mathcal{T}^\Psi[i,j,\kappa]\mathcal{T}^\Psi[j,\kappa^\prime,i] \left.\mathcal{M}^\mathrm{GR}(i,A,j,\kappa)\vphantom{\mathcal{M}^{R^3}}\right|^{\epsilon_{\kappa}\to \epsilon_n}_{\Sigma(\kappa)} \times \left.\mathcal{M}^\mathrm{GR}(1,\ldots,i,j,\ldots, \kappa^\prime)\vphantom{\mathcal{A}^{R^3}}\right|^{\epsilon_{\kappa^\prime}\to \epsilon_n}_{\Sigma(\kappa^\prime)},
\end{multline}

\begin{multline}
    \mathcal{M}_n^{R^2} \to
\mathcal{T}^\Psi[i,j,\kappa]\mathcal{T}^W[j,\kappa^\prime,i] \left.\mathcal{M}^{R^2}(i,A,j,\kappa)\right|^{\epsilon_{\kappa}\to \epsilon_n}_{\Sigma(\kappa)} \times \left.\mathcal{M}^{R^2}(1,\ldots,i,j,\ldots, \kappa^\prime)\right|^{\epsilon_{\kappa^\prime}\to \epsilon_n}_{\Sigma(\kappa^\prime)},
\end{multline}

\begin{multline}
    \mathcal{M}^{R^3}_n \to
\mathcal{T}^W[i,j,\kappa]\mathcal{T}^W[j,\kappa^\prime,i] \left.\mathcal{M}^{R^3}(i,A,j,\kappa)\right|^{\epsilon_{\kappa}\to \epsilon_n}_{\Sigma(\kappa)} \times  \left.\mathcal{M}^{R^3}(1,\ldots,i,j,\ldots, \kappa^\prime)\right|^{\epsilon_{\kappa^\prime}\to \epsilon_n}_{\Sigma(\kappa^\prime)}.
\end{multline}


We could of course compare the action of the transmuting operators before and after the splitting, but the details are not very illuminating.


\section{Smooth splitting and hidden zeros}

In this section, we extend our analysis to the 3-split process of \cite{Cachazo:2021wsz}. The three fixed punctures, denoted by $\mathcal{I}=\{i,j,k\}$, with $i<j<k$, divide the full set of $n$-external particles into three disjoint subsets, which we label $A$, $B$, and $C$.

Under the conditions
\begin{equation}
    s_{a,b} =   s_{b,c} =  s_{c,a} =0 \qquad \forall \, a\in A,\, b\in B\,, c\in C\,,
    \label{eq:main.3-split}
\end{equation} 
the integration measure becomes
\begin{multline}\label{eq:measure3-split}
    \dd\mu_n  \to \Delta(i,j,k)^{2}\left[ \Delta(i,j,\kappa_A)^{-2} \dd\mu_{\mathrm{A}}(i, j, A; \kappa_A) \right] \\ \times \left[ \Delta(j,k,\kappa_B)^{-2} \dd\mu_{\mathrm{B}}(j, k, B; \kappa_B)\right] 
    \left[ \Delta(k,i,\kappa_C)^{-2}\dd\mu_{\mathrm{B}}(k, i, C; \kappa_C)\right],
\end{multline}
with
\begin{equation}
\begin{array}{ccc}\displaystyle 
    \dd\mu_A = \Delta(i,j,\kappa_A)^{2}\,{\prod_{a\in A}}\dd z_a\,\delta\left(S_a\right),& & \displaystyle  S_{a}=\frac{s_{a,i}}{z_{a,i}}+\frac{s_{a,j}}{z_{a,j}}+\frac{s_{a,\kappa_A}}{z_{a,\kappa_A}}+\sum_{\tilde{a}\in A\setminus\{a\}}\frac{s_{a,\tilde{a}}}{z_{a,\tilde{a}}},\\
   \displaystyle  \dd\mu_B = \Delta(j,k,\kappa_B)^{2}\,{\prod_{b\in B}}\dd z_b\,\delta\left(S_b\right),& &\displaystyle  S_{b}=\frac{s_{b,\kappa_B}}{z_{b,\kappa_B}}+\frac{s_{b,j}}{z_{b,j}}+\frac{s_{b,k}}{z_{b,k}}+\sum_{\tilde{b}\in B\setminus\{b\}}\frac{s_{b,\tilde{b}}}{z_{b,\tilde{b}}}, \\
    \displaystyle  \dd\mu_A = \Delta(k,i,\kappa_C)^{2}\,{\prod_{c\in C}}\dd z_c\,\delta\left(S_c\right), & &\displaystyle  S_{c}=\frac{s_{b,i}}{z_{b,i}}+\frac{s_{b,\kappa_C}}{z_{b,\kappa_C}}+\frac{s_{b,k}}{z_{b,k}}+\sum_{\tilde{c}\in C\setminus\{c\}}\frac{s_{c,\tilde{c}}}{z_{c,\tilde{c}}},
\end{array}
\end{equation}
where $\{\kappa_A,\kappa_B,\kappa_C\}$ respectively replace the legs $\{k,i,j\}$, and represent off-shell states with
\begin{subequations} \begin{align}
p_{\kappa_A} & =-p_i-p_j-\sum_{a\in A} p_a, \\
p_{\kappa_B} & =-p_j-p_k-\sum_{b\in B} p_b, \\
p_{\kappa_C} & =-p_k-p_i-\sum_{c\in C} p_c.
\end{align} \end{subequations}

\noindent It is also straightforward to check that the Parke--Taylor factor can be rewritten as
\begin{multline}\label{eq:PT3-split}    
 \mathrm{PT}(1,2,\dots,n) =  
    \Delta(i,j,k)^{-1}\left[ \Delta(i,j,\kappa_A){\rm PT}(i,A,j,\kappa_A)\right] \\ \times\left[ \Delta(j,k,\kappa_B){\rm PT}(j,B,k,\kappa_B)\right]
    \left[\Delta(k,i,\kappa_C){\rm PT}(k,C,i,\kappa_C\right]. 
\end{multline}

With this basic ingredients, it can be shown \cite{Cachazo:2021wsz}, for instance, that the tree level amplitude for the bi-adjoint scalars has a 3-split form given by
\begin{align}
	\mathcal{A}^{\phi^3}(1,2,\dots,n)\to&\, \mathcal{J}^{\phi^3}(i,A,j,\kappa_A)\times \mathcal{J}^{\phi^3}(j,B,k,\kappa_B)\times \mathcal{J}^{\phi^3}(k,C,i,\kappa_C), \\
	\to&\, \mathcal{A}^{\phi^3}(i,A,j,\kappa_A)\Big|_{\Sigma(\kappa_A)}\times \mathcal{A}^{\phi^3}(j,B,k,\kappa_B)\Big|_{\Sigma(\kappa_B)}\times \mathcal{A}^{\phi^3}(k,C,i,\kappa_C)\Big|_{\Sigma(\kappa_C)}. \nonumber
\end{align}
For the second line, we have simply applied the results of section \ref{sec:on-shellrep}. In the particular case in which $A$ has just one element, i.e. $A=\{ a \}$, the above equation leads to
\begin{equation}\label{eq:HzeroBA}
	\mathcal{A}^{\phi^3}(1,2,\dots,n)\to 
	\left( \frac{1}{s_{a,i}} + \frac{1}{s_{a,j}} \right)
	\times \mathcal{J}^{\phi^3}(j,B,k,\kappa_B)\times \mathcal{J}^{\phi^3}(k,C,i,\kappa_C).
\end{equation}
When $-(s_{a,i}+s_{a,j})=s_{a,\kappa_A}=s_{a,k}=0$, the right hand side of \eqref{eq:HzeroBA} vanishes,  establishing the connection between the 3-split behavior and hidden zeros of the amplitude \cite{Arkani-Hamed:2023swr}.

Now, let us focus on the integrand $W_{\{1,2,\dots,n\}}$.
Since the sets $A$, $B$, and $C$ are mutually disjoint, we impose the 3-split conditions
\begin{equation}
    \epsilon_{a}\cdot p_b = \epsilon_{a}\cdot p_c = 0, \quad \epsilon_{b}\cdot p_a = \epsilon_{b}\cdot p_c= 0, \quad\epsilon_{c}\cdot p_a = \epsilon_{c}\cdot p_b=0, \quad \forall \, a\in A,\, b\in B\,, c\in C\,.
    \label{eq:main2.3-split}
\end{equation}
Then, it follows that $W_{\{1,2,\dots,n\}}$ splits as
\begin{equation}
W_{\{1,2,\dots,n\}} \to \prod_{a\in A}\left(-\sum_{\substack{\tilde{a}\in A\cup\{i,j,\kappa_A\}\\
(\tilde{a}\neq a)
}
}\frac{\epsilon_{a}\cdot p_{\tilde{a}}}{z_{a,\tilde{a}}}\right) 
\prod_{b\in  B}\left(-\sum_{\substack{\tilde{b}\in B\cup\{j,k,\kappa_B\}\\
(\tilde{b}\neq b)
}
}\frac{\epsilon_{b}\cdot p_{\tilde{b}}}{z_{b,\tilde{b}}}\right)
W_{C\cup\{k,i,\kappa_C\}},
\label{eq:3splitW}
\end{equation}

Equation \eqref{eq:3splitW}, together with \eqref{eq:measure3-split} and \eqref{eq:PT3-split}, imply the 3-split of the $(DF)^2$ amplitude, expressed as
\begin{multline}\label{eq:DF2-splitting}
{\cal{A}}_n^{(DF)^2}(1,2,\dots,n) \to {\cal{J}}^{(DF)^2+\phi^3}(i^\phi,A,j^\phi,\kappa_A^\phi) \\ \times 
{\cal{J}}^{(DF)^2+\phi^3}(j^\phi,B,k^\phi,\kappa_B^\phi) \times 
{\cal{J}}_{\mu}^{(DF)^2}(k,C,i,\kappa_C)\epsilon_j^\mu.
\end{multline}

\noindent In the special case $A = \{ a \} $, the 3-split behavior of \eqref{eq:DF2-splitting} is recast as
\begin{multline}
\mathcal{A}_n^{(DF)^2}(1,2,\dots,n) \to 
\left(
\frac{ \epsilon_a\cdot p_i }{s_{a,i}}-\frac{ \epsilon_a\cdot p_j }{s_{a,j}}
\right) \\
\times 
\mathcal{J}^{(DF)^2+\phi^3}(j^\phi,B,k^\phi,\kappa_B^\phi) \times 
\mathcal{J}_{\mu}^{(DF)^2}(k,C,i,\kappa_C)\epsilon_j^\mu.
\label{eq:DF2-3splitspecial}
\end{multline}
In order to uncover the hidden zero, we impose $s_{a,\kappa_A}=s_{a,k}=0$ and $ \epsilon_a\cdot p_{\kappa_A}=\epsilon_a\cdot p_{k}=0$. It is then straightforward to show that the first term on the right hand side of \eqref{eq:DF2-3splitspecial} vanishes.

Naturally, this analysis extends to the $R^3$ theory. By imposing the splitting conditions \eqref{eq:main2.3-split} on both $W_{\{1,\ldots,n\}}$ factors, the 3-split behavior is immediately revealed, either as
\begin{equation}
\mathcal{M}_n^{R^3} \to \mathcal{J}^{R^3+\phi^3}(i^\phi,A,j^\phi,\kappa_A^\phi) \times \mathcal{J}^{R^3+\phi^3}(j^\phi,B,k^\phi,\kappa_B^\phi) \times \mathcal{J}_{\mu\nu}^{R^3}(k,C,i,\kappa_C)\epsilon_j^\mu\epsilon_j^\nu,
\label{eq:R33split}
\end{equation}
or as
\begin{multline}
\mathcal{M}_n^{R^3} \to \mathcal{J}^{R^3+\phi^3}(i^\phi,A,j^\phi,\kappa_A^\phi) \\ \times \mathcal{J}_{\mu}^{R^3+(DF)^2}(j^g,B,k^g,\kappa_B^g)\epsilon_i^\mu \times \mathcal{J}_{\nu}^{R^3+(DF)^2}(k^g,C,i^g,\kappa_C^g)\epsilon_j^\nu,
\label{eq:R33splitalt}
\end{multline}
where the superscript $g$ denotes gluon states from the  $(DF)^2$ theory.

Analogously to the $(DF)^2$ case with $A = \{ a \} $, the additional conditions $s_{a,\kappa_A}=s_{a,k}=0$ and $ \epsilon_a\cdot p_{\kappa_A}=\epsilon_a\cdot p_{k}=0$ lead to the vanishing the amplitude.


\section{Final remarks}

In this paper, we have extended the universal 2-split behavior of tree-level scattering amplitudes of particles and strings found in \cite{Cao:2024gln,Cao:2024qpp} to amplitudes in higher-derivative theories. This is achieved via a simple analysis of the split behavior of $W_{\{1,2,\dots,n\}}$, cf. equation \eqref{eq:Wn}, which is the key ingredient in  the CHY representations of the $(DF)^2$ gauge theory and the $R^2$ gravity amplitudes \cite{Azevedo:2017lkz}. As expected, the splitting of $W_{\{1,2,\dots,n\}}$ is analogous to the splitting of the reduced Pfaffian. Therefore, gluon and graviton amplitudes in those theories display analogous splitting behaviors.

Our main result, inspired by a recent work by one of the authors \cite{Gomez:2025tqx} and discussed in section \ref{sec:on-shellrep}, is the recasting of the 2-split behavior in terms of amplitudes rather than CHY currents, i.e. directly in terms of \textit{physical} quantities. This result extends to all theories previously analyzed under the 2-split behavior. Within the higher derivative models, we further simplify their analysis with the introduction of new transmuting operators and discuss their 3-split behavior in line with reference \cite{Cachazo:2021wsz}.

We hope this new on-shell characterization of the 2-split behavior will help with the understanding of the so-called hidden zeros of scattering amplitudes. Furthermore, it would be interesting to investigate to what extent it could be extended to loop-level or even to massive theories. The latter, in particular, have been notoriously elusive to  the CHY framework, which might indicate that other methods would be better suited to the investigation of their possible 2-split behavior. 

We would like to emphasize that our results are consistent with the 2-split structure of bosonic string theory. For example, as shown in \cite{Azevedo:2018dgo} by one of the authors and collaborators, the scattering amplitude of $n$-gluons in the open bosonic string can be schematically written  in a double-copy form,
\begin{equation}\label{eq:dcbosonic}
  {\cal A}_n^{\rm bos.}(1,2,\ldots, n) = (\text {Z-theory}) \otimes ((DF)^2 + YM),  
\end{equation}
where the tensor product is defined through the field theory KLT (Kawai–Lewellen–Tye) representation \cite{Kawai:1985xq}. The Lagrangian for the $(DF)^2 + \mathrm{YM}$ theory is given by
\begin{align}
\mathcal{L}_{(DF)^2+YM} &=  \frac{1}{2}(D_{\mu} F^{\mu \nu\,I})^2  + \frac{g}{3} \, F^3+ \frac{1}{2}(D_{\mu} \varphi^{\alpha})^2  + \frac{g}{2}  \,  C^{\alpha IJ}  \varphi^{ \alpha}   F_{\mu \nu}^I F^{\mu \nu\,J }  \nonumber\\ 
&
+  \frac{g}{3!}  \, d^{\alpha \beta \gamma}   \varphi^{ \alpha}  \varphi^{ \beta} \varphi^{ \gamma}
- \frac{1}{2}m^2 (\varphi ^\alpha)^2 -\frac{1}{4}m^2(F^{\mu \nu\,I})^2 , 
\end{align}
with
\begin{equation}
m^2=-\frac{1}{\alpha'}\, ,
\end{equation}
  where $\alpha'$ sets the fundamental length scale of the string.

On the other hand, the 2-split behavior for the scattering of $n$-gluons in the open bosonic string was derived in \cite{Cao:2024qpp}, and is expressed as
\begin{equation}\label{eq:2splitbosonic}
  {\cal A}_n^{\rm bos.}(1,2,\ldots, n) \to   {\cal J}^{\rm bos. + color}(i^\phi,j^\phi,A, \kappa^\phi)  \times {\cal J}_\mu^{\rm bos.}(1,\ldots, i,j, \ldots,n-1, \kappa') \, \epsilon_n^\mu \, ,
\end{equation}
where the superscript “bos.+color” denotes the mixed current with three $\phi^3$ scalars.

Taken together, the double copy structure in equation \eqref{eq:dcbosonic} and the 2-split behavior in equation \eqref{eq:2splitbosonic} clearly show that the pure Yang-Mills 2-split representation given in equation \eqref{eq:YMsplit} is exactly recovered in the field theory limit ($\alpha'\to 0$).
Moreover, in the tensionless limit ( $\alpha' \to \infty$), one precisely obtains the result presented in equation \eqref{eq:DF2-splitting_}. In contrast, recent results \cite{Zhang:2026dcm} for the $F^3$ theory display an unconventional 2-split behavior, which nevertheless remains fully compatible with the bosonic string structure given in equation \eqref{eq:2splitbosonic}. Since their analysis focuses on a finite contribution in the $\alpha'$-expansion of the open bosonic string, the resulting splitting behavior receives contributions from multiple combinations of distinct types of currents. This interplay leads to the unusual 2-split structure.

\paragraph*{Acknowledgments}
HG and RLJ are supported by the GA\v{C}R grant 25-16244S from the Czech Science Foundation. The work of TA was partially supported by the European Structural and Investment Funds and the Czech Ministry of Education, Youth and Sports (project FORTE CZ.02.01.01/00/22\_008/0004632), through its research mobility program.

\appendix

\section{Independent Mandelstam invariants}\label{IMI}

In this appendix, we describe an alternative method to the one presented in Section~\ref{sec.kinematic}. The construction is inspired by the ideas developed by one of the authors in \cite{Gomez:2025tqx}. 
  
 Prior to formulating a general statement about the appropriate choice of independent kinematic variables, we find it instructive to consider two illustrative examples: the five-point cases in the BA and $(DF)^2$ theories.
    
Let us recall that a five-point amplitude involves $\tfrac{5(5-3)}{2} = 5$ independent kinematic invariants, and  the complete set of $\frac{5(5-1)}{2} = 10$ Mandelstam variables $s_{i,j}$ can be organized into a symmetric $5 \times 5$ matrix. 
According to the prescription introduced in \cite{Gomez:2025tqx}, one can rearrange this matrix such that the first, second, and the last columns correspond to the three punctures fixed by the ${\rm SL}(2,\mathbb{C})$ gauge symmetry. In the specific BA example considered  in section~\ref{sec:example.On-shell}, ${\cal I}=\{i,j,k\} = \{1,3,5\}$,  while the intermediate columns correspond to the labels represented by the sets $A = \{2\}$ and $B = \{4\}$.

Following the procedure outlined in \cite{Gomez:2025tqx}, we can choose the five independent kinematic invariants for the current on the left-hand side of equation \eqref{eq:2splitBA5}, namely ${\cal J}^{\phi^3}(i,A,j,\kappa) = {\cal J}^{\phi^3}(1,2,3,\kappa)$, as the set represented by the blue-shaded entries in the left matrix in~\eqref{matrix5a}, 
{\small
\begin{equation}\label{matrix5a}
\begin{tikzpicture}[scale=0.95,transform shape]
\matrix [matrix of math nodes]
[left delimiter=(,right delimiter=)] (m){
           0 & s_{1,3} &  s_{1,4}  &  s_{1,2} & s_{1,5} \\
           {} & 0 & s_{3,4} & s_{3,2} & s_{3,5} \\ 
            {} & {} & 0 & s_{4,2} & s_{4,5} \\ 
            {} & {} & {} & 0  & s_{2,5}  \\
              {} & {} & {} & {}  & 0  \\};
                 \draw (m-3-5.south west) -- (m-3-5.south east)[opacity=1,thick];
                  \draw (m-3-5.north west) -- (m-3-5.north east)[opacity=1,thick];
                  \draw (m-3-5.south west) -- (m-3-5.north west)[opacity=1,thick];
                    \draw (m-3-5.south east) -- (m-3-5.north east)[opacity=1,thick];
        \draw[ fill=blue, opacity=0.13] (m-3-4.south west) rectangle (m-1-4.north east);
         \draw[ fill=blue, opacity=0.13] (m-2-3.north west) rectangle (m-1-4.north west);
        \end{tikzpicture}\, ,\qquad\qquad
        \begin{tikzpicture}[scale=0.95,transform shape]
\matrix [matrix of math nodes,left delimiter=(,right delimiter=)] (m){
           0 & s_{1,3} &  s_{1,2}  &  s_{1,4} & s_{1,5} \\
           {} & 0 & s_{3,2} & s_{3,4} & s_{3,5} \\ 
            {} & {} & 0 & s_{2,4} & s_{2,5} \\ 
            {} & {} & {} & 0  & s_{4,5}  \\
              {} & {} & {} & {}  & 0  \\};
                 \draw (m-3-5.south west) -- (m-3-5.south east)[opacity=1,thick];
                  \draw (m-3-5.north west) -- (m-3-5.north east)[opacity=1,thick];
                  \draw (m-3-5.south west) -- (m-3-5.north west)[opacity=1,thick];
                    \draw (m-3-5.south east) -- (m-3-5.north east)[opacity=1,thick];
        \draw[ fill=blue, opacity=0.13] (m-3-4.south west) rectangle (m-1-4.north east);
         \draw[ fill=blue, opacity=0.13] (m-2-3.north west) rectangle (m-1-4.north west);
        \end{tikzpicture}\, , 
        \end{equation}
        }\noindent 
along with the ``mass" of the off-shell leg, $p_{\kappa}^2=(p_1+p_2+p_3)^2=(p_4+p_5)^2=s_{1,2,3}=s_{4,5}$. Notice that the middle columns of this matrix are arranged as $(B,A) = (\{4\},\{2\})$. This choice is essential for maintaining the independence of the five Mandelstam invariants. If one instead adopted the opposite ordering, $(A,B) = (\{2\},\{4\})$, the invariants would become linearly dependent through the relation $s_{1,4} + s_{2,4} + s_{3,4} + s_{4,5} = 0$. We denote the set of these highlighted Mandelstam variables as ${\bf K}^L_5=\{ s_{1,4} \,, s_{1,2}\, , s_{2,3} \, , s_{2,4} \}$.

For the right current in equation \eqref{eq:2splitBA5}, ${\cal J}^{\phi^3}(1,\ldots,i,j,\ldots,\kappa^\prime) = {\cal J}^{\phi^3}(1,3,4,\kappa^\prime)$, the five independent kinematic invariants are indicated by the blue-shaded entries in the right matrix in~\eqref{matrix5a}, together with the off-shell leg ``mass" $p_{\kappa}^{_{'}2}=(p_1 + p_3+p_4)^2=(p_2 + p_5)^2 =s_{1,3,4}= s_{2,5}$. Note that the middle columns of this matrix follow the opposite ordering to those in the left matrix,   i.e.  $(A,B) = (\{2\},\{4\})$. We denote the set of these highlighted kinematic variables as ${\bf K}^R_5=\{ s_{1,2} \,, s_{1,4}\, , s_{2,4} \, , s_{3,4} \}$.

Having determined appropriate sets of independent Mandelstam invariants, we evaluate the amplitudes $\mathcal{A}_4^{\phi^3}(1,2,3,\kappa)$ and $\mathcal{A}_4^{\phi^3}(1,3,4,\kappa^\prime)$ using, respectively, the sets ${\bf K}^L_5 \cup \{s_{4,5}\}$ and ${\bf K}^R_5 \cup \{s_{2,5}\}$. These computations are carried out via the bi-adjoint Berends–Giele recursion \cite{Mafra:2016ltu}, leading to
\begin{align}
\left.\mathcal{A}_4^{\phi^3}(1,2,3,\kappa)\right|_{{\bf K}^L_5}&= \left.\left[ \lim_{s_{1,2,3}\rightarrow 0} s_{1,2,3}\,{\cal J}_{{\rm BG}}^{\phi^3}(1,2,3,\kappa)\right]\right|_{{\bf K}^L_5}= \frac{1}{s_{1,2}}+\frac{1}{s_{2,3}}, \\
\left.\mathcal{A}_4^{\phi^3}(1,3,4,\kappa^\prime)\right|_{{\bf K}^R_5}&= \left.\left[\lim_{s_{1,3,4}\rightarrow 0} s_{1,3,4}\,{\cal J}_{{\rm BG}}^{\phi^3}(1,3,4,\kappa^\prime)\right]\right|_{{\bf K}^R_5}= \frac{1}{s_{3,4}}+\frac{1}{-s_{1,4}-s_{3,4}},  
\end{align}
which precisely reproduces the result found in equations \eqref{eq:CHY5pts1} and \eqref{eq:CHY5pts2}, as desired.

\subsection{$(DF)^2$ at five-point, 2-split}

As a second five-point example, we examine the same configuration, ${\cal I}=\{i,j,k\} = \{1,3,5\}$, with $A = \{2\}$ and $B = \{4\}$, applied to the amplitude $\mathcal{A}_5^{(DF)^2}(1,2,3,4,5)$. Within this setup, the corresponding two-split conditions are given by,
\begin{equation}\label{Split5pt}
s_{2,4}=0, \qquad \epsilon_{2}\cdot p_4=0, \qquad p_2\cdot \epsilon_{b^\prime}=0,\quad {b^\prime} \in B \cup {\cal I}.  	
\end{equation}
With this configuration, the 2-split takes the form, 
\begin{equation}
\mathcal{A}_5^{(DF)^2}(1,2,3,4,5) \,\,
^{\underrightarrow{\quad}}\,\,
 {\cal J}^{(DF)^2+\phi^3}(1^\phi,2,3^\phi,\kappa^\phi)\times {\cal J}^{(DF)^2}_\mu(1,3,4,\kappa^\prime) \, \epsilon^\mu_5.		
\end{equation}
where $p_\kappa=-p_1-p_2-p_3=p_4+p_5$, and $p'_{\kappa}=-p_1-p_3-p_4=p_2+p_5$.
Rather than computing the amputated currents directly, we proceed to evaluate the four-point amplitudes $\mathcal{A}_4^{(DF)^2+\phi^3}(1^\phi,2,3^\phi,\kappa^\phi)$ and $\mathcal{A}_4^{(DF)^2}(1,3,4,\kappa^\prime)$, employing the independent kinematic variables specified in~\eqref{matrix5a}, i.e. ${\bf K}^L_5 \cup \{s_{4,5}\}$ and ${\bf K}^R_5 \cup \{s_{2,5}\}$, respectively. Thus, using the expressions presented in eq:~\eqref{eq:4pt-DF2}, we arrive at
\begin{align}\label{DF25pMs}
&
\left.\mathcal{A}_4^{(DF)^2+\phi^3}(1^\phi,2,3^\phi,\kappa^\phi)\right|_{{\bf K}^L_5}= 
{\cal T}^{W}[1,3,\kappa]\, 
\left.\mathcal{A}_4^{(DF)^2}(1,2,3,\kappa)\right|_{{\bf K}^L_5}=
\left( \frac{ p_{1} {\cdot} \epsilon_2}{s_{1,2}} - \frac{ p_3 {\cdot} \epsilon_2 }{s_{2,3}} \right) 
, \nonumber\\
&
\left.\mathcal{A}_4^{(DF)^2}(1,3,4,\kappa^\prime)\right|_{{\bf K}^R_5 
}= \frac{ (s_{1,4}+s_{3,4})^2s^2_{3,4}}{ s_{1,4} } 
\left(\frac{ p_3 {\cdot} \epsilon_1 }{s_{1,4}+s_{3,4}} -\frac{ p_{34} {\cdot} \epsilon_1}{s_{3,4}} \right)
\left( \frac{ p_{3} {\cdot} \epsilon_4}{s_{3,4}} - \frac{ p_{13} {\cdot} \epsilon_4 }{s_{1,4}+s_{3,4}} \right)
\nonumber\\
&
\qquad\qquad\qquad\qquad\qquad\,\,\,\,
\left( -\frac{ p_{1} {\cdot} \epsilon_3}{s_{1,4}+s_{3,4}} - \frac{ p_4 {\cdot} \epsilon_3 }{s_{3,4}} \right)
\left( \frac{ p_{13} {\cdot} \epsilon_{\kappa^\prime}}{s_{1,4}+s_{3,4}} - \frac{ p_1 {\cdot} \epsilon_{\kappa^\prime} }{s_{3,4}} \right),   \nonumber\\
\end{align}
which shows perfect agreement with ${\cal J}^{(DF)^2+\phi^3}(1^\phi,2,3^\phi,\kappa^\phi)$ and ${\cal J}_\mu^{(DF)^2}(1,3,4,\kappa^\prime)\epsilon_{\kappa^\prime}^\mu$. If the split conditions specified in equation \eqref{Split5pt} are applied to equation \eqref{DF25pMs}, we obtain the expected result, which reads
\begin{align}
&
\mathcal{A}_5^{(DF)^2}(1,2,3,4,5) \,\,
^{\underrightarrow{\quad}}\,\,
{\cal T}^{W}[1,3,\kappa]\,
 \mathcal{A}_4^{(DF)^2}(1,2,3,\kappa)\Big|_{{\bf K}^L_5} \times \mathcal{A}^{(DF)^2}_4(1,3,4,\kappa^\prime) \Big|_{{\bf K}^L_5}^{\epsilon_{\kappa^\prime}\rightarrow \epsilon_5}\nonumber\\	
 &
 \qquad\qquad\quad\qquad\quad\quad
 =
 \frac{ s_{4,5}^2s^2_{3,4}}{ s_{1,4} } 
\left( \frac{ p_{1} {\cdot} \epsilon_2}{s_{1,2}} - \frac{ p_3 {\cdot} \epsilon_2 }{s_{2,3}} \right) 
\left( \frac{ p_{5} {\cdot} \epsilon_1}{s_{3,4}} -\frac{ p_3 {\cdot} \epsilon_1 }{s_{4,5}} \right)
\left( \frac{ p_{3} {\cdot} \epsilon_4}{s_{3,4}} - \frac{ p_{5} {\cdot} \epsilon_4 }{s_{4,5}} \right)
\nonumber\\
&
\qquad\qquad\qquad\qquad\quad\quad
\left( \frac{ p_{1} {\cdot} \epsilon_3}{s_{4,5}} - \frac{ p_4 {\cdot} \epsilon_3 }{s_{3,4}} \right)
\left( \frac{ p_{4} {\cdot} \epsilon_{5}}{s_{4,5}} - \frac{ p_1 {\cdot} \epsilon_{5} }{s_{3,4}} \right). 	
\end{align}

These examples illustrate how the method developed in \cite{Gomez:2025tqx} can be straightforwardly applied to the 2-split computation, allowing amputated currents to be extracted directly from amplitudes and thereby simplifying the overall procedure. Having examined and analyzed these cases, we are now in a position to formulate a general statement.

\subsection{General independent kinematic variables sets}


Before proceeding to the general case, it is worth noting that, in this work, we leave the number of spacetime dimensions unspecified. Instead, we consider the kinematic space defined solely in terms of the Mandelstam invariants $s_{a,b} = (p_a + p_b)^2=2\,p_a\cdot p_b$. These variables are not all independent, as momentum conservation imposes the constraints $\sum_{b=1}^n s_{a,b} = 0$. Consequently, a scattering process involving $n$ massless external particles contains
\begin{equation}
\frac{n(n-1)}{2} - n = \frac{n(n-3)}{2}
\end{equation}
independent kinematic invariants.

Building on the insights gained from the previous examples and the approach proposed in \cite{Gomez:2025tqx}, we now outline the general framework.
To compute the amputated currents resulting from the 2-split procedure, which schematically takes the form
\begin{equation}
\mathcal{A}(1,2,\dots,n) \,\,
^{\underrightarrow{\quad}}\,\,	
	{\cal J}^{L}(i,A,j,\kappa)\times {\cal J}^{R}(1,\ldots,i,j,\ldots,\kappa^\prime),
\end{equation} 
we choose   suitable sets of independent Mandelstam invariants corresponding to the upper-triangular entries of the following symmetric $s_{a,b}$ matrices, respectively, 
\begin{equation}\label{matrix5}
\begin{tikzpicture}[scale=0.95,transform shape]
\matrix [matrix of math nodes]
[left delimiter=(,right delimiter=)] (m){
           i & j &  B  &  A & k \\};
        \end{tikzpicture}\, ,\qquad\qquad
        \begin{tikzpicture}[scale=0.95,transform shape]
\matrix [matrix of math nodes,left delimiter=(,right delimiter=)] (m){
           i & j &  A  &  B & k \\};
        \end{tikzpicture}\, , 
        \end{equation}
as illustrated in the matrix representation of equation \eqref{matrix5}.

More precisely, these matrices yield $\tfrac{n(n-3)}{2} - 1$ independent kinematic variables in the following way:
from their upper-triangular part, we remove the columns labeled $(i,j,k)$ and, in addition, the entry $s_{j,j+1}$ in the left matrix and $s_{j,i+1}$ in the right one.\footnote{Recall that the sets $A$ and $B$ are given by $A=\{i+1,\ldots,j-1 \}$ and $B=\{j+1,\ldots,n-1,1,\ldots,i-1 \}$, where $A\cup B\cup \{i,j,n\}=\{1,2,\ldots,n\}$.} Consequently, the remaining entries define the sets ${\bf K}^L_n$ and ${\bf K}^R_n$, each containing $\tfrac{n(n-3)}{2} - 1$ independent Mandelstam invariants that specify the kinematic data used to evaluate the corresponding amputated currents.
To complete the full sets of independent kinematic variables,   i.e.  $\tfrac{n(n-3)}{2}$ invariants,
 we must also include the ``masses'' of the off-shell legs, namely
\begin{align}
p_\kappa^2 &=(p_B+p_n)^2=(p_{j+1}+ \cdots + p_{i-1} + p_n)^2 = s_{j+1,\ldots,i-1,n}, \\
p_{\kappa}^{_{'}2} &= (p_A+p_n)^2=(p_{i+1}  + \cdots + p_{j-1} + p_n)^2 = s_{i+1,\ldots,j-1,n}.
\end{align}
Accordingly, the complete and consistent sets of $\tfrac{n(n-3)}{2}$ independent Mandelstam invariants are given by
\begin{align}
{\bf K}^L_n \cup \{ s_{j+1,\ldots,i-1,n}  \},
\qquad
{\bf K}^R_n \cup \{ s_{i+1,\ldots,j-1,n} \}.
\end{align}
Thus, as in the preceding five-point examples, we can use these two sets to evaluate the amputated currents ${\cal J}^L(i,A,j,\kappa)$ and ${\cal J}^R(1,\ldots,i,j,\ldots,\kappa^\prime)$ directly from their amplitude counterparts,
\begin{align}
	{\cal J}^L(i,A,j,\kappa)&=\mathcal{A}^L(i,A,j,\kappa)\Big|_{{\bf K}^L_n } ,\\
    {\cal J}^R(1,\ldots,i,j,\ldots,\kappa^\prime)&=\mathcal{A}^R(1,\ldots,i,j,\ldots,\kappa^\prime)\Big|_{{\bf K}^R_n }. 
\end{align}
Here it is important to recall that, in the amplitude sector, one must impose the on-shell conditions,
$p_\kappa^2  = s_{j+1,\ldots,i-1,n}=0$ and $p_{\kappa}^{_{'} 2} = s_{i+1,\ldots,j-1,n}=0$.

Finally, we can now express the 2-split relation in a compact and consistent form, as a product of lower-point amplitudes,
\begin{equation}
\mathcal{A}(1,2,\dots,n) \,\,
^{\underrightarrow{\quad}}\,\,	\mathcal{A}^L(i,A,j,\kappa)\Big|_{{\bf K}^L_n } \times \mathcal{A}^R(1,\ldots,i,j,\ldots,\kappa^\prime)\Big|_{{\bf K}^R_n }.
\end{equation}

\bibliographystyle{unsrtnat}
\bibliography{references}

\end{document}